\documentclass[aps,pra,letterpaper,superscriptaddress,twocolumn,showpacs,floatfix,10pt]{revtex4-1}

\usepackage{amsfonts}
\usepackage{amsmath}
\usepackage{amssymb}
\usepackage{graphicx}
\usepackage{epstopdf}
\usepackage{epsfig}
\usepackage{color}
\usepackage{mathtools}
\usepackage{hyperref}
\usepackage{tabularx}
\newcommand{\be}{\begin{eqnarray}}
\newcommand{\ee}{\end{eqnarray}}

\newcommand{\bfr}{{\bf r}}

\newcommand{\bfk}{{\bf k}}

\newcommand{\wbe}{\begin{widetext}}
\newcommand{\wee}{\end{widetext}}

\begin{document}

\title{Quantum Degenerate Majorana Surface Zero Modes in Two-Dimensional Space}

\author{Ching-Yu Huang}
\affiliation{ Physics Division, National Center for Theoretical Sciences,  Hsinchu, Taiwan}

\author {Yen-Ting Lin}
\affiliation{  Institute for Theory of Statistical Physics, RWTH Aachen, 52056 Aachen}

\author { Hao Lee}  
\affiliation{  Physics Department, National Tsing Hua University, Hsinchu, Taiwan}

\author {Daw-Wei Wang}
\affiliation{ Physics Division, National Center for Theoretical Sciences,  Hsinchu, Taiwan}
\affiliation{  Physics Department, National Tsing Hua University, Hsinchu, Taiwan}

\vfill

\begin{abstract}
We investigate the topological properties of spin polarized fermionic polar molecules loaded in a multi-layer structure with the electric dipole moment polarized to the normal direction.
 When polar molecules are paired by attractive inter-layer interaction, unpaired Majorana fermions can be macroscopically generated in the top and bottom layers in dilute density regime. 
 We show that the resulting topological state is effectively composed by a bundle of 1D Kitaev ladders labeled by in-plane momenta ${\bfk_\|}$ and ${-\bfk_\|}$, and hence belongs to BDI class characterized by the winding number $\mathbb{Z}$, protected by the time reversal symmetry. 
 The Majorana surface modes exhibit a flatband at zero energy, fully gapped from Bogoliubov excitations in the bulk, and hence becomes an idea system to investigate the interaction effects on quantum degenerate Majorana fermions. 
 We further show that additional interference fringes can be identified as a signature of such 2D Majorana surface modes in the time-of-flight experiment.
\end{abstract}

\date{\today}
\maketitle

\section{Introduction}
A Majorana fermion (MF) is known to be its own antiparticle \cite{MF}, as a hypothesis in theoretical particle physics. 
In condensed matter systems, a MF can appear as a localized edge state and reflect the topological property in the bulk of system. 
It is known that topological system can be classified into intrinsic topological order or symmetry-protected topological (SPT) order: the latter is robust against small perturbations for a given on-site symmetries  \cite{classification}. 
Ground states of nontrivial SPT phases cannot be continuously connected to trivial product states without either closing the gap or breaking the protecting symmetry \cite{Bosonic_SPT}. 

One of the mostly studied topological phase is proposed by Kitaev \cite{Kitaev_1D} for one-dimensional (1D) $p$-wave superconductor. 
The Majorana zero mode (MZM) of such system may be applied for quantum computation through braiding, and certain experimental signature has been proposed in an ordinary $s$-wave superconducting wire with strong spin-orbital coupling through proximity effect 
\cite{Proximity_MF, S_Heterostructures, tunable_S_device, Spin_Singlet_SC, S_SC_Heterostructures, Signatures_of_MF,  Zero_bias_peaks, ac_Josephson, Anomalous_Zero_Bias_Conductance, QAHE_SC} 
or even in systems of ultracold atoms 
\cite{MZM_Fermionic_Cold_Atoms, MF_Cold_Atom_Quantum_Wires, 2D_Spin_Orbit, s_Wave_SF, Novel_p_wave}. 
However non-ambiguous evidence is still lacking probably due to the poor signal-to-noise ratio for a localized MZM. 
Recently, some extensions of 1D Kitaev model by including inter-chain tunneling \cite{multichains}, dimerization  \cite{dimerize}, and long-ranged pairing  \cite{long_range_hopping_and_pairing, long_range_interaction} have also been proposed.

In the systems of quantum gases, such inter-site pairing between fermions can be provided by long-ranged dipolar interactions between polar molecules \cite{Ni, Zwierlein_1,Zwierlein_2, Ketterle, Weidemuller_1,Weidemuller_2} or neutral atoms with a large magnetic moment \cite{Vernac, Lev, Ferlaino}. 
Since the inelastic coupling between polar molecules can be well suppressed by aligning dipole moments normal to the two-dimensional (2D) plane with a strong transverse confinement \cite{Bohn1,BaranovM,YeM}, many exotic ground states have been predicted in a 2D layer \cite{Taylor, Cooper2009, Gora, Pikovski2010, Sun, Miyakawa, DasSarmaFL, BaranovSieb, ZinnerDW, Parish, Babadi, Lu1, Lu2, Giorgini} or in a bi/multi-layer structure  \cite{Pikovski2010, ZinnerDW, Ronen, FabioDW, Potter,Babadi}.

\begin{figure}[htb!]
 \centering
\includegraphics[width=0.5\textwidth]{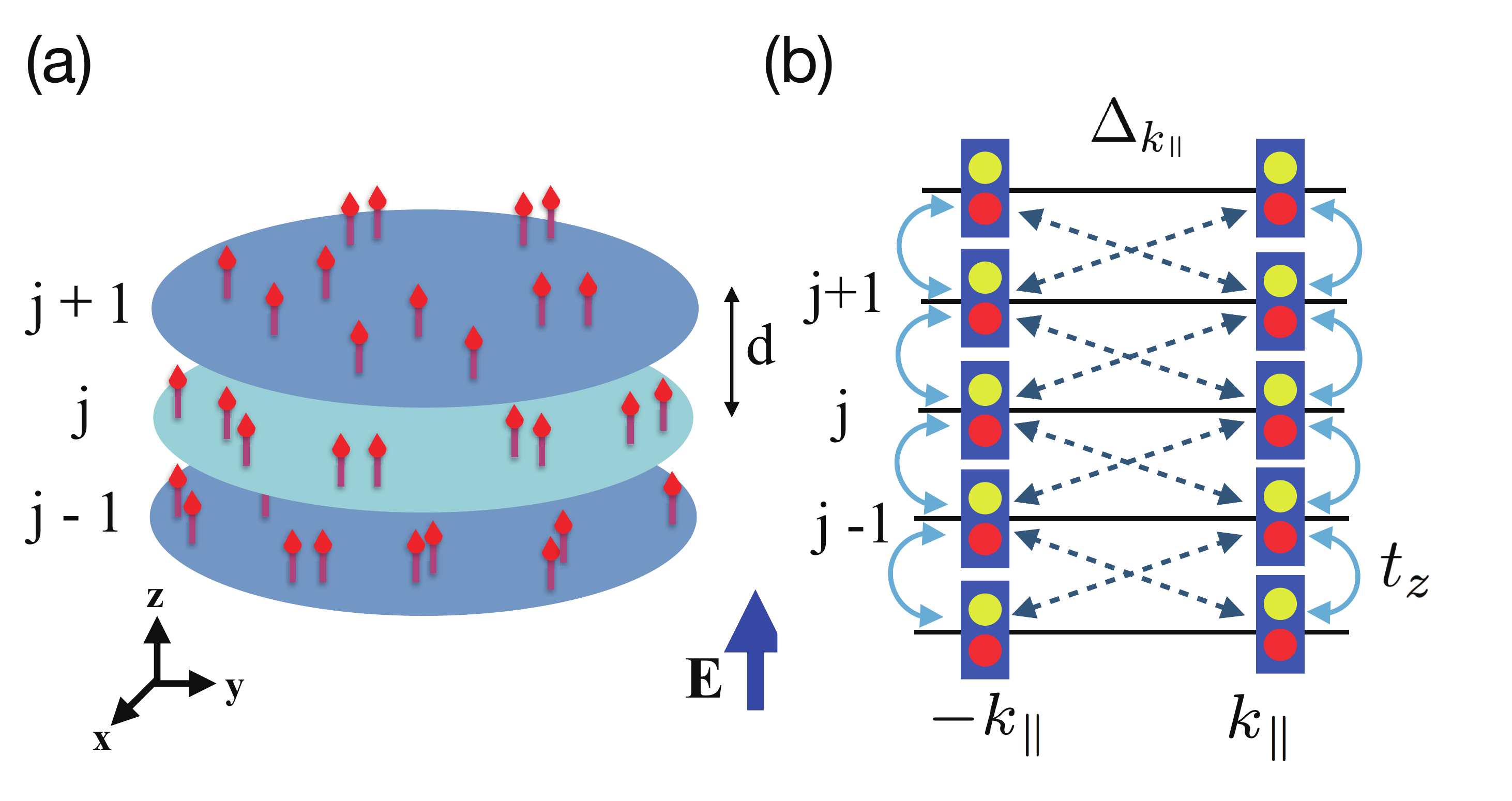}
\caption{
(a) Multi-layer structure for identical fermionic polar molecules loaded into a deep 1D optical lattice. 
The electric dipole moment (arrows) is aligned to the normal ($z$) direction by the external field. 
(b) Effective Kitaev ladder in momentum space (see Eq.~(\ref{Hbcs})), where fermions (rectangular boxes) in each "site" can be decomposed into two MFs (circles, see the text).
 Two unpaired MF can be found at the end of ladders if Eq.~(\ref{phase boundary2}) is satisfied.
 Double-sided arrows of dashed/sold lines indicate inter-layer pairing/hopping respectively (for simplicity, here we just show the configuration of a special case, where only pairs between neighboring layers are considered, see Sec.~\ref{special case} ). }
\label{fig:mutilayer}
\end{figure}

In this paper, we proposed that macroscopically degenerate Majorana Surface Zero Modes (MSZMs) can exist in a 2D continuous space, where spin polarized fermionic polar molecules are paired and hence perform a superfluid in a stack of multi-layer structure (Fig.~\ref{fig:mutilayer}(a)) through their dipolar interaction. 
Within the BCS theory and Landau-Fermi liquid theory (both are well-justified in our 3D system), we show that the multi-layer system becomes equivalent to an ensemble of 1D Kitaev ladders along the normal direction (see Fig.~\ref{fig:mutilayer}(b)), labelled by in-plane momenta $(\bfk_\|,-\bfk_\|)$. 
This is, each subsector is equivalent to the BdG Hamiltonian of 1D Kitaev ladders. 
The resulting topological state is actually belong to BDI class, characterized by the $\mathbb{Z}$ index as the winding number. 
The associated topological properties are confirmed by both the analytical calculation as well as the numerical calculation of entanglement spectrum/entropy when the long-range pairing is included. 
These unpaired MSZMs are protected by the time reversal symmetry as well as the superfluid pairing gap, showing a localized wavefunction along the normal direction of layers ($z$) and a mobile flat energy band in the in-plane dictions ($x-y$) inside the superfluid gap. 
From experimental point of view, such topological state can be observed through the additional short period interference fringes, which are resulted from macroscopically occupied MSZMs in the top and bottom layers. 
Our results therefore  suggest an interesting system to investigate quantum many-body properties of non-Abelian anyons in a 2D continuous space.

Our paper is organized as follows: 
in Sec.~\ref{system}, we setup a system which are loaded into a stack of 2D layers with spin polarized fermionic polar molecules, and derive the effective model Hamiltonian within the meanfield approximation.
In Sec.~\ref{short-range}, we investigate the topological properties and Majorana Surface Zero Modes of a special case, where only the nearest pairing is considered. 
We then extend our calculation in Sec.~\ref{long-range} toward a more general case when the long-ranged pairing (due to dipolar interaction) is included. 
In Sec.~\ref{stability} we show why these MFs are stable against single particle perturbation due to the time-reversal symmetry. 
Finally we discuss some experimentally related issues in Sec.~\ref{experiment} and conclude our paper in Sec.~\ref{conclusion}. 
In Appendix ~\ref{appA}, we show how the unpaired MFs can be analytically obtained from the Bogoliubov-de Gennes (BdG) equation. 
We then explicitly calculate the winding number in Appendix~\ref{appB}.

\section{System Hamiltonian}
\label{system}

\subsection{Full Hamiltonian and Meanfield Approximations}

The system setup is illustrated in Fig.~\ref{fig:mutilayer}(a), where spin polarized fermionic polar molecules are loaded into a stack of 2D layers with interlayer spacing $d$ and tunneling amplitude $t_z$. 
The electric dipole moment is polarized and perpendicular to the layer by the external electric field, leading to an attractive inter-layer interaction and a repulsive intra-layer interaction between molecules. 
As a result, the full system Hamiltonian can be expressed as following in the real space coordinates:
\wbe
\be
H &=&-t_z\sum_{j=1}^{L-1}\int d\bfr_\|\left[\psi_j(\bfr_\|)^\dagger\psi_{j+1}(\bfr_\|)+{\rm h.c.}\right]
+\sum_{j=1}^{L}\int d\bfr_\| \psi_j(\bfr)^\dagger\left[\frac{-\nabla^2}{2m}
+U_{\rm trap}(\bfr_\|)-\mu\right]\psi_j(\bfr_\|)
\nonumber\\
&&+\frac{1}{2}\sum_{j,j'=1}^L\int d\bfr_\|\int d\bfr_\|'
V_d(\bfr_\|-\bfr'_\|,jd-j'd)\psi_j(\bfr_\|)^\dagger\psi_{j'}(\bfr_\|')^\dagger\psi_{j'}(\bfr_\|')\psi_j(\bfr_\|),
\label{H_original}
\ee
\wee
where $m$ and $\mu$ are the mass and chemical potential of polar molecules.
$\psi_j(\bfr_\|)$ is the field operator for the layer index $j$ and the in-plane coordinate, $\bfr_\|$. 
$V_d(\bfr_\|,z)$ is dipole-dipole interaction between polar molecules.
Here for simplicity, we assume the layer width is much smaller than inter-layer distance and can be neglected. 
$U_{\rm trap}(\bfr_\|)$ is the in-plane trapping potential, and $L\gg 1$ is the number of total layers.

Since the multi-layer structure described above is actually a quasi-3D system, and therefore ordinary meanfield approximations are well justified. 
In order to highlight the possibility of SPT order in such 3D systems, we consider the parameter regime when the averaged in-plane density is in the dilute regime, i.e., the averaged inter-particle distance between molecules in the same layer is larger than the inter-layer distance. 
It is then reasonable to expect that the effects of in-plane repulsive interaction ($\sim |\bfr_\| |^{-3}$) are relatively weaker than both the in-plane kinetic energy and the effects of inter-layer attractive interaction (see Sec.~\ref{experiment} below for more details). 
As a result, the attractive inter-layer interaction pairs up polar molecules in different layers as a superfluid phase within the BCS theory
 By contrast, the repulsive interaction between molecules in the same layer is a relatively weak effect, renormalizing the effective mass and chemical potential within the Landau-Fermi liquid theory \cite{Pikovski2010, DasSarma}. 
Throughout this paper, we will NOT discuss the situation when the in-plane repulsion becomes relevant, which appears only in the high density regime.  
We will briefly discuss its possible effects in Sec.~\ref{stability}.

\subsection{Effective Hamiltonian of Kitaev ladders}

In the rest of this paper, we consider that the trapping potential $U_{\rm trap}(\bfr_\|)$ is assumed to be shallow enough so that we could neglect it and apply periodic boundary condition in the in-plane $(x-y)$ direction.
We will discuss the situation of finite size effect in Sec.~\ref{stability} and show it should be instant to the stability of the topological properties. 
As a result, within the mean field approximations mentioned above, an effective Bogoliubov-de Gennes (BdG) Hamiltonians can be easily derived to be: $H_{\rm BdG}=\sum_{\bfk_\|}H_{\bfk_\|}$, where
\begin{eqnarray}
H_{\bfk_\|}
&=&\sum_{j=1}^{L}\left(\frac{\bfk_{\parallel}^2}{2m^\ast}-\mu^\ast\right)c_{\bfk_{\parallel},j }^{\dagger}c_{\bfk_{\parallel},j }   \notag \\
&-&t_z^\ast\sum_{j=1}^{L-1}\left[c_{\bfk_{\parallel},j}^{\dagger}c_{\bfk_{\parallel},j+1 }^{}+{\rm h.c.}\right]    \notag \\
&-&\sum_{j\neq j',jj'=1}^L
\Delta^{(|j'-j|)}_{\bfk_\|}\left[c_{\bfk_{\parallel},j}^{\dagger}c_{-\bfk_{\parallel},j'}^{\dagger}+{\rm h.c.}\right].
\label{Hbcs}
\end{eqnarray}
Here $c_{\bfk_\|,j}=\frac{1}{\sqrt{\Omega}}\int d\bfr_\|\psi_j(\bfr_\|)\,e^{-i\bfk_\|\cdot\bfr_\|}$ is field operator of the $j$th layer and the in-plane momentum, $\bfk_\|$. 
$\Omega$ is area of the 2D layer; $\mu^\ast$, $m^\ast$, and $t_z^\ast$ are renormalized chemical potential, molecule mass, and interlayer tunnelling amplitude respectively. 
$\Delta^{(|j'-j|)}_{\bfk_\|}$ is gap function between the $j$th and $j'$th layers. 

One can see that if we define 
\begin{eqnarray}
H_{\bfk_\|}^{\rm ladd}\equiv H_{\bfk_\|}+H_{-\bfk_\|}.
\label{H_ladd1}
\end{eqnarray}
$H_{\bfk_\|}^{\rm ladd}$ is nothing but a two-leg Kitaev ladder in {\it momentum space}, where the topological properties (if exist) are obviously protected by time-reversal, particle-hole, and chiral symmetries at the same time, and hence belongs to BDI class of Altland-Zirnbauer classification~\cite{classification}. 
Here the time-reversal symmetry is effectively protected, because polar molecules can be initially prepared in the same hyperfine state with a polarized spin state~\cite{hyperfine}.
And, their exchange interaction is known so weak that spin relaxation is almost frozen during the experimental holding time.

\section{MSZMs for the Nearest-Neighbor Pairing} 
\label{short-range}

In order to investigate the possible topological properties of the Kitaev ladder in momentum space, we would start from a specific model, where superfluid pairing appears only between nearest neighboring layers, i.e. $\Delta^{(|j'-j|)}_{\bfk_\|}=0$ for $|j-j'|\geq 2$. 
Results including longer range interaction will be discussed in the next section.

\subsection{Exact Solution of a Special Case}
\label{special case}

Firstly, we show the exact solution of a special case and then calculate the winding number for a general situation. 
We consider a special situation when $\bfk_{\parallel}^2/2m^\ast-\mu^\ast=0$ and $\Delta_{\bfk_\|}\equiv \Delta^{(1)}_{\bfk_\|}=t_z^\ast$. 
This special choices makes the ladder Hamiltonian, $H_{\bfk_\|}^{\rm ladd}$, to have a single energy scale only.

We then define MFs in momentum space to be: 
\begin{equation}
\gamma_{\bfk_\|,j}^{(1)}\equiv c_{\bfk_\|,j}^{}+c_{-\bfk_\|,j}^\dagger \quad  \gamma_{\bfk_\|,j}^{(2)}\equiv -i(c_{\bfk_\|,j}^{}-c_{-\bfk_\|,j}^\dagger ), 
\end{equation}
with the anti-commutation relations,  $\{\gamma^{(a)}_{\bfk_\|,j},\gamma^{(b)}_{\bfk_\|',j'}\}=2\delta_{j,j'}\delta_{a,b}\delta_{\bfk_\|,-\bfk_\|'}$ ($a,b=1,2$). 
After some algebra, we obtain (see Appendix~\ref{appA})
\begin{equation}
H_{\bfk_\|}^{\rm ladd}=it_z^\ast\sum_{j=1}^{L-1}\left(\gamma_{\bfk_\|,j}^{(2)}\gamma_{-\bfk_\|,j+1}^{(1)}+\gamma_{-\bfk_\|,j}^{(2)}\gamma_{\bfk_\|,j+1}^{(1)}\right),  
\label{H_ladd2}
\end{equation}
where the ladder Hamiltonian becomes two decoupled chains with two MFs missed at each end (see Fig.~\ref{fig:mutilayer}(b)): $\gamma_{\pm\bfk_\|,1}^{(1)}$ for the bottom layer ($j=1$), and $\gamma_{\pm\bfk_\|,L}^{(2)}$ for the top layer ($j=L$). 
One can easily construct two "real" MFs in each layer by defining 
$ \bar{\gamma}_{\pm,\bfk_\|}^{(a)}=(\pm1)^{\frac{1}{2}}({\gamma}_{\bfk_\|}^{(a)} \pm {\gamma}_{-\bfk_\|}^{(a)})$, which is its own anti-particle: $\left(\bar{\gamma}_{\pm,\bfk_\|}^{(a)}\right)^\dagger=\bar{\gamma}_{\pm,\bfk_\|}^{(a)}$  $a=1,2$. 

It is easy to show that, even for ladders with $k_\|^2/2m\neq \mu$ or $\Delta_{\bfk_\|}\neq t^\ast_z$, we could still diagonalize $H_{\bfk_\|}^{\rm ladd}$ by introducing new MFs, $\tilde{\gamma}_{\bfk_\|,l}^{(a)}$, which is a linearly combination of $\gamma_{\bfk_\|,j}^{(a)}$ ($a=1,2$), and satisfies  $\{\tilde{\gamma}_{\bfk_\|,l}^{(a)},\tilde{\gamma}_{\bfk_\|',l'}^{(b)}\}=2\delta_{l,l'}\delta_{a,b}\delta_{\bfk_\|,-\bfk_\|'}$. 
Localized MFs, $\tilde{\gamma}_{\pm\bfk_\|,j=1/L}^{(1)}$, near the bottom/top layer can be still retained when some condition is satisfied (see below). 
It is just an extension of 1D Kitaev chain to two-ladder case in the momentum space.

\subsection{General Condition for MSZMs}
\label{general condition}

The topological property of our Kitaev ladder in momentum space can be further investigated by calculating the winding number with a periodic boundary condition along the normal ($z$) direction. 
After applying Fourier transform on the Hamiltonian (Eq.~(\ref{Hbcs})) along the $z$ direction.
For simplicity, here we just consider the nearest neighbouring pairing only, although it can be in principle applied to a more general situation. 
Defining Nambu spinor, $C^{}_{k_{\parallel},k_z}=[c^{}_{k_{\parallel},k_z},c_{-k_{\parallel},-k_z}^{\dagger}]^T $, we obtain,
\begin{equation}
H_{BdG} = \frac{1}{2}\sum_{k_{\parallel}}\sum_{k_z}C^{\dagger}_{k_{\parallel},k_z}
\left[\begin{array}{cc}                
\xi_\bfk  & \tilde{\Delta}^\dag_\bfk \\
\tilde{\Delta}_\bfk & -\xi_\bfk 
\end{array}
\right]                
C_{k_{\parallel},k_z},
\label{H_C}
\end{equation}
where $\xi_\bfk\equiv {\bfk_\|^2}/ {2m^\ast}-\mu^\ast-2t_z^\ast\cos(k_z d)$. 
Note that we define 3D momentum $\bfk\equiv (\bfk_\|,k_z)$ and 3D gap function $\tilde{\Delta}_\bfk\equiv 2i\Delta_{\bfk_\|}\sin(k_zd)$ for the convenience of later discussion. 
The Bogoliubov excitation energies are given by 
\be
E_\bfk^{(\pm)} &=& \pm \sqrt{\xi_\bfk^2+|\tilde{\Delta}_\bfk|^2}
\nonumber\\
&=& \pm \sqrt{\xi_\bfk^2+4|\Delta_{\bfk_\|} |^2\sin^2(k_zd)}.
\label{excitation}
\ee
The gap is closed at $k_z=0$ or $\pi/d$, and  $\xi_\bfk\equiv  {\bfk_\|^2}/{2m^\ast}-\mu^\ast-2t^\ast_z\cos(k_z d)= {\bfk_\|^2}/{2m^\ast}\mp\mu^\ast-2t^\ast_z=0$.  

To calculate the winding number of the Bloch state wavefunction, we first apply a unitary transformation on the BdG Hamiltonian to make it off-diagonal,
\be
H_{BdG} &=& \frac{i}{4}\sum_{\bfk}
\Gamma^{\dagger}_{\bfk}  \left[ 
  \begin{array}{cc}   
    0 & v_{\bfk} \cr      
    v_{\bfk}^{\dagger} & 0 \cr  
  \end{array}
\right]    \Gamma_{\bfk},
\label{H_gamma}
\ee
with $v_\bfk\equiv\xi_\bfk+\frac{1}{2}( \tilde{\Delta}^\ast_\bfk- \tilde{\Delta}_\bfk)=\xi_\bfk-2i\Delta_{\bfk_\|}\sin(k_z d) \equiv R(\bfk)\mathrm{e}^{i\theta (\bfk)}$ and 
$\Gamma^{}_{\bfk}\equiv [c^{}_{\bfk}+ c_{-\bfk}^{\dagger},  -ic^{}_{\bfk}+i c_{-\bfk}^{\dagger}]^T $.  

The winding number can be calculated as follows (see Appendix~\ref{appB} for details) : 
\begin{eqnarray}
W_{\bfk_\|} 
& \equiv &\int_{-\pi/d}^{\pi/d}\frac{dk_z}{2\pi}\partial_{k_z}\theta_{\bfk_\|,k_z} 
\\
&=&\frac{J_{-}}{2}\oint \frac{dz}{2\pi i} \frac{ z^2+2 J_{+} z+ 1}{J_{1}J_{2}\prod_{j=1}^4(z-Z_j)]} 
\nonumber
\label{winding number calculation}
\end{eqnarray}
where $\mu_{\bfk_\|}^\ast \equiv \mu^\ast- {\bfk_\|^2}/{2m^\ast}$, $J_{1}=\frac{t_z^\ast+\Delta_{\bfk_\|}}{\mu_{\bfk_\|}^\ast}$, $J_{2}=\frac{t_z^\ast-\Delta_{\bfk_\|}}{\mu_{\bfk_\|}^\ast}$, and $J_{\pm}=J_{1}\pm J_{2}$. 
Four poles of the integrand are given as
$Z_{1,2}=\frac{-1 \pm \sqrt{1-4 J_1 J_2} }{2 J_1}$ and $Z_{3,4}=\frac{-1 \pm \sqrt{1-4 J_1 J_2} }{2 J_2}$ respectively. 
One can find that $W_{\bfk_\|}$ become nonzero (i.e. topological non-trivial) only when the following condition is satisfied: 
\begin{equation}
\left|\mu^\ast-\frac{\bfk_\|^2}{2m^\ast}\right|< 2t_{z}^\ast,
\label{phase boundary2}
\end{equation}
which is independent of the gap function as expected. 
This condition fails as $\bfk_{\parallel}^2/2m^\ast\pm 2t_z=\mu^\ast$, exactly when the Bogoliubov excitation energy becomes gapless at $k_z=0$ and $\pi/d$ (see Eq.~(\ref{excitation}).

From Eq.~(\ref{phase boundary2}),  the Majorana surface modes appears when the in-plane momentum is within a certain range: 
\begin{eqnarray}
\label{k_range}
k_{\rm Min} < |\bfk_\| | < k_{\rm Max}, 
\end{eqnarray}
where $k_{\rm Min}\equiv \sqrt{2m^\ast({\rm Max}(0,\mu^\ast-2t_z^\ast))}$ and $k_{\rm Max}\equiv \sqrt{2m^\ast(\mu^\ast+2t_z^\ast)}$ respectively. 

Throughout this paper, we will consider the situation when $t_z^\ast>\mu^\ast/2$ {\it only}, so that $k_{\rm Min}=0$ and ladders of {\it all} in-plane momenta for $|\bfk_\| |<k_{\rm Max}$ are topologically nontrivial. 
(Note that, $k_{\rm Max}$ is also the largest Fermi momentum around the Fermi sea at $k_z=0$.) 
Under this condition, the Majorana modes occupy the whole in-plane Fermi sea in the top and bottom layers. 
From the energy point of view, these unpaired MFs stay in the middle of the superfluid gap.  
As for the situation when $\mu^\ast > 2 t_z^\ast$,  only those states with in-plane momenta, $k_{\rm Min} < |\bfk_\| | < k_{\rm Max}$, are possible states with topological non-trivial properties. 
However, as we will show later, the topological properties of such a case may become unstable against disorder potential. 
We will therefore not emphasize this case in the rest of this paper.

\section{MSZMs for a Long-Ranged Paring}
\label{long-range}

In this section, we will include the long-ranged pairing order parameter into the Kitaev ladder (see Eq.~(\ref{Hbcs})(\ref{H_ladd1})), where the pairing amplitude is expected to be, $\Delta^{(|j-j'|)}_{\bfk_\|}=\Delta^{(1)}_{\bfk_\|}\times |j-j'|^{-3}$, due to the dipolar interaction. 
The major goal is to investigate if the longer-ranged pairing will make such topological state unstable or to change the boundary of phase diagram, compared to the nearest neighboring pairing in the last section.  
In principle, we could repeat the calculation of winding number and investigate the effects of longer-range pairing. However, the analytical calculation becomes much more complicated and hence less intuitive.
Therefore, we will show a numerical results based on entanglement spectrum and entropy as well as the exact diagonalization methods. 
Both of them confirm the topological properties even with dipolar interaction.

\begin{figure}[htb!]
 \centering
\includegraphics[width=0.5\textwidth]{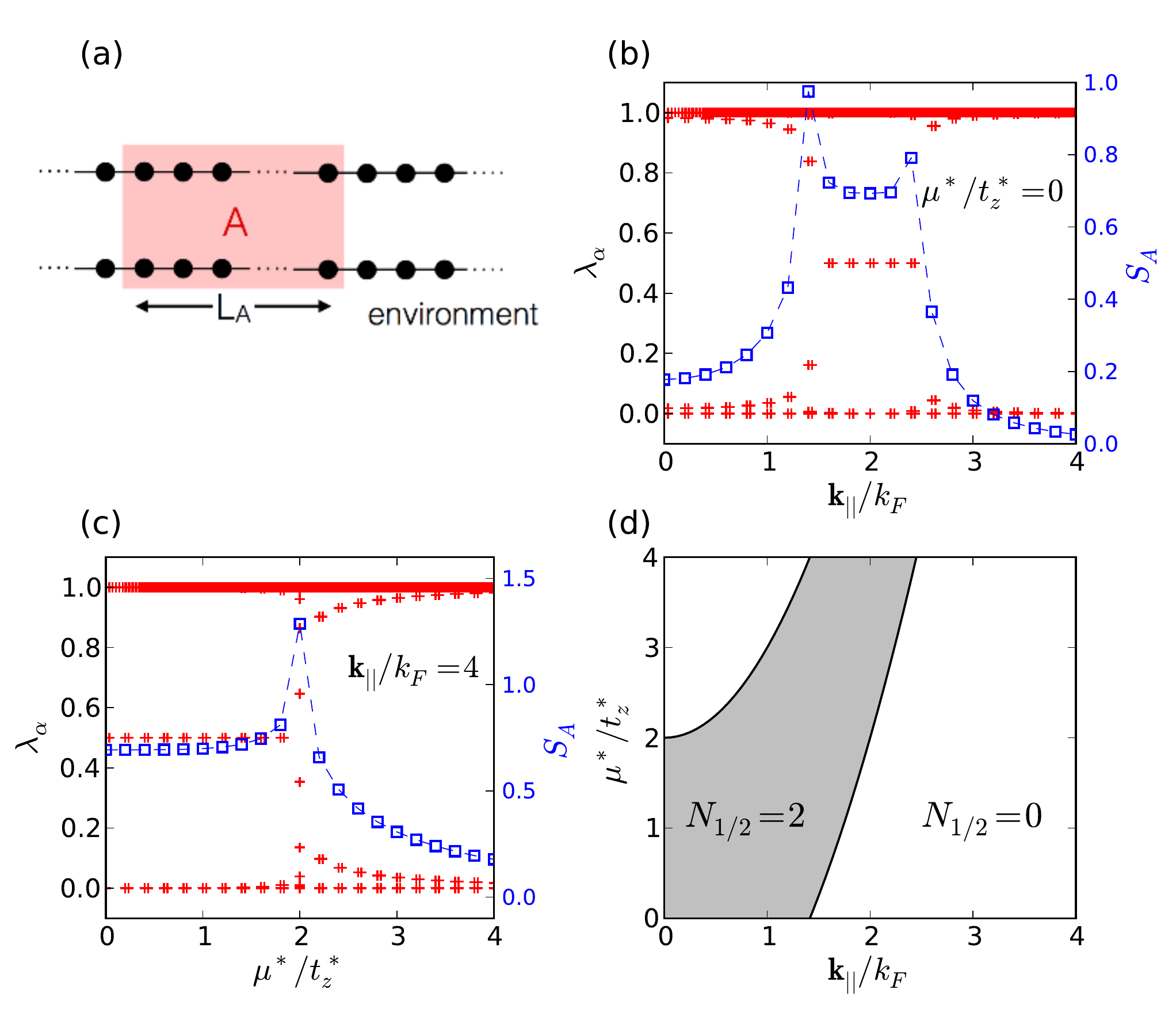}
\caption{  
(a) The whole system is divided into finite subsystem A with $L_A$ sites and environment. 
Here we consider $L \gg L_A\gg 1$. 
(b) shows the entanglement spectrum, $\lambda_{\alpha}$ (red plus), and the entanglement entropy, $S_A$ (blue square) as a function of $\bfk_\| /k_F$ for $\mu^\ast=4 t_z^\ast$.
Here $k_F\equiv \sqrt{2m^\ast \mu^\ast}$.
(c) is the same as (b) but as a  function of $\mu^\ast/t_z^\ast$ for $\bfk_\|=0$.
(d) Phase diagram of the topological regime calculated from entanglement spectrum. 
$N_{1/2}$ is the number of the degeneracy in the entanglement spectrum at $\lambda_\alpha=1/2$ for the subsystem $A$. 
This is also indicating the number of MFs at the edge. 
The upper and lower boundaries are exactly the same as $k_{\rm Min}$ and $k_{\rm Max}$ calculated from Eq.~(\ref{phase boundary2}).
} 
\label{fig2}
\end{figure}

\subsection{Full Numerical Calculation of Entanglement Spectrum and Entanglement Entropy}

A topological phase can be also characterized by the quantum entanglement between the subsystem and the environment~\cite{EN_TO_1,EN_TO_2, ES_FQHE,  ES_SPT_1,ES_SPT_2}.
In short, given a ground state wave function $|\Psi\rangle$, one can calculate the reduced density matrix, $\rho_A$, for a subsystem $A$ by tracing over the environment~(see Fig.~\ref{fig2}(a)). 
The eigenvalues $\lambda_\alpha$ of the reduced density matrix is so-called ``entanglement spectrum"~\cite{ES_FQHE}, which carries nonlocal information and has been applied for calculating Berry phase and zero-energy edge states~\cite{berry_phase}. 
For example, the degeneracy of entanglement spectrum has recently implemented to characterize the topological properties for some 2D quantum Hall states~\cite{ES_FQHE} and for some 1D SPT phases~\cite{ES_SPT_1,ES_SPT_2}.

For 1D Kitaev model, $\lambda_\alpha$ is given by the eigenvalues of the block Green's function matrix, i.e.,  $G_{\bfk_\|,i,j}\equiv\langle c^{}_{\bfk_\|,i}c^\dagger_{\bfk_\|,j} \rangle$ with the layer indices $i$ and $j$ inside the subsystem $A$. 
In Ref.~\cite{berry_phase, opes_1,opes_2}, it has been shown that the zero energy mode of1D Kitaev model corresponds to the degeneracy of $\lambda_{\alpha} =1/2$ in the entanglement spectrum, i.e., the pair of zero modes at two ends of Kitaev chain contribute the maximal entanglement between the subsystem $A$ and environment. 
Besides of entanglement spectrum, a topological phase transition can be also identified by the discontinuity of the entanglement entropy of the subsystem (given by $S_A = - {\rm Tr} [ \rho_A \log \rho_A]$) after tracing out the environment.

It has been shown that the entanglement spectrum $\lambda_{\alpha}$ of the subsystem $A$ can be obtained by diagonalizing the entire Green's function matrix $G_{{\bfk_\|},m,n}$~\cite{berry_phase,opes_1,opes_2}, where $m$ and $n$ are restricted in the subsystem $A$ (along z-direction) with wave number ${\bfk_\|}$. 
To calculate the Green's function, we first express the full mean field Hamiltonian in another form (see Eq.~(\ref{H_C})): $H_{BdG} = \frac{1}{2}\sum_{\bfk_\|, k_z}  C_{{\bfk_\|},k_z}^{\dagger} [ \mathbf{R}(\bfk) \cdot  \vec{ \sigma} ]   C_{{\bfk_\|},k_z}  $, where $\vec{\sigma}= (\sigma_x, \sigma_y, \sigma_z )$ are Pauli matrices and $R(\bfk)\equiv (0,  \tilde{\Delta}_\bfk ,\xi_\bfk)$. 
As a result, the Green's function matrix defined in real space lattice sites (i.e. layer), 
\begin{equation}
G_{{\bfk_\|},m,n} = \frac{1}{L} \sum_{k_z \in BZ} e^{i k_z (m-n)d} G_{\bfk },
\end{equation}
where  $d$ is inter-layer distance,  $L$ is the total number sites of $z$-direction, and $k_z$ takes values in the first Brillouin zone 
It is worth mentioning that  $G_{\bfk}$ being a $2 \times 2$ matrix,
$G_{\bfk} = \frac{1}{2} \left(1 + \frac{ {\mathbf{R}}(\bfk) \cdot \vec{\sigma } } {| \mathbf{R}(\bfk) |}\right) $.

In our case, we numerically diagonalize the block's Green's function for subsystem $A$ with a finite size, e.g., $L_A=100$ and $L=200$, of the ladder as shown in  Fig.~\ref{fig2}(a). 
The numerical results converges and is independent of the choices of subsystem in the thermal dynamic limit, $L\gg L_A\gg 1$.
In Figs.~\ref{fig2}(b) and (c), we show the entanglement spectra $\lambda_{\alpha}$ and entanglement entropy $S_A$ obtained by calculating the  Green's function from the effective Hamiltonian, Eq.~(\ref{Hbcs}). 
The topological regime for $\lambda_\alpha=1/2$ appears through a topological phase transition at which, the entanglement entropy diverges. 
In Fig.~\ref{fig2}(d), we show the regime of topological non-trvial phase ($\lambda_\alpha=1/2$), which is exactly the same regime as defined by Eq.~(\ref{phase boundary2}).  
These numerical results agree with the analytic calculation of the winding number, confirming the topological properties in our current system.

\begin{figure}[htb!]
\centering
\includegraphics[width=0.45\textwidth]{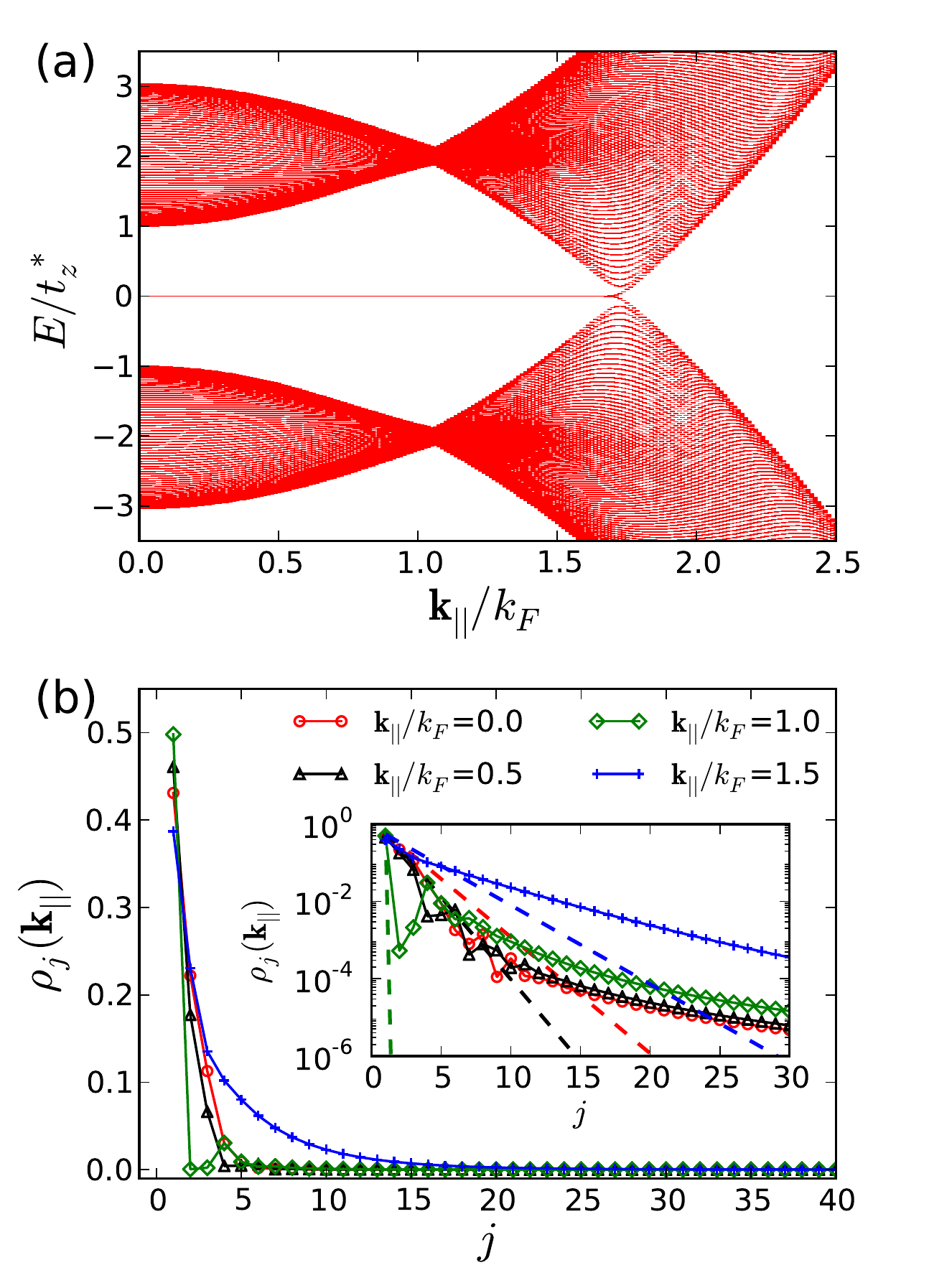}
\caption{
(a) Single particle band structure of ladder Hamiltonian (See Eq.~(\ref{H_ladd1})) obtained by exactly diagonalizing L=100 layers with a long-ranged (dipolar) pairing for $\mu^\ast=\Delta_{\bfk_\|}^{(1)}=t_z^\ast$. 
(b) Particle density distribution $\rho_j(\bfk_\|)$ in the real space for the Majorana zero modes for different in-plane momentum $\bfk_\|$ with $\mu^\ast=\Delta_{\bfk_\|}^{(1)}=t_z^\ast$. 
Inset: Semi-log plot for the same data. The dashed lines are results for nearest neighbouring pairing only. }
\label{fig:flatband long range}
\end{figure}
%

\subsection{Flatband Structure and Edge State Wavefunction}

In our numerical calculation above, the topological properties of the Kitaev ladder (and hence the multi-layer structure in our original proposal) have been confirmed in a periodic boundary condition along the normal ($z$) direction. 
However, the localized edge state wavefunction as well as the possible zero energy modes still have to be obtained within an open boundary condition.

In this section, we perform a full exact diagonalization on the meanfield Hamiltonian in Eq.~(\ref{H_ladd1}), where a long-ranged order parameter (pairing) is included with an open boundary condition.
 In Fig.~\ref{fig:flatband long range}(a), we show the obtained single particle energy as a function of the in-plane momentum $|\bfk_\| |=k_x$ ( we choose $k_y=0$) for $\mu^\ast=\Delta_{\bfk_\|}^{(1)}=t_z^\ast$ with finite layer number $L=100$. 
 The continuous energy bands above and below zero energy is the Bogoliubov excitation spectrum, 
while a new zero energy flat band appears between zero momentum and $k_{\rm Max}=\sqrt{2m^\ast (\mu^\ast+2t_z^\ast)}=\sqrt{3}k_F$, where $k_F\equiv \sqrt{2m^\ast\mu^\ast}= \sqrt{2m^\ast t_z^\ast}$. 
 In Fig.~\ref{fig:flatband long range}(a), due to finite size effect, the energy spectra split near gap closed point.

From Bogoliubov excitation energies in Eq.~\ref{excitation}, it is easy to show that the gap is closed at $k_x=k_{\rm Max}$, and $k_y=k_z=0$. 
The presence of such zero energy flatband structure indicates that the Majorana surface zero modes do exist in our current 3D system, while its topological properties are classified to be BDI class as Kitaev chain. 
We note that, such quantum degenerate MSZMs have overlapping wavefunction in the 2D surface, and therefore become a very interesting systems for investigating the many-body physics of non Abelian particles. 
We will then discuss its stability in details in the next section.

When $\mu^\ast> 2t_z^\ast$, however, the physical properties is changed: there will be a ring structure in the momentum space, where the Kitaev ladder becomes topological as $k_{\rm Min}<|\bfk_\| |<k_{\rm Max}$, see  Eq.~(\ref{phase boundary2}). 
In addition to the gapless ring as $|\bfk_\| |=k_{\rm Max}$, the Bogoliubov excitation energy gap also closes at $|\bfk_\| |=k_{\rm Min}$. 
However, different from previous case, the appearance of these new gapless ring may cause additional coupling between Majorana fermions and the Bogoliubov particles, making the topological phases unstable against local disorder potential. 
We will discuss this issue in the next section.

Besides of energy spectrum, in Fig.~\ref{fig:flatband long range}(b), we also show the density distribution, 
$\rho_j(\bfk_\|)\equiv \langle\psi^{0}_{\bfk_\|}|c^\dagger_{\bfk_\|,j}c^{}_{\bfk_\|,j}|\psi^{0}_{\bfk_\|}\rangle$, for the Majorana zero modes alone the normal direction (z-direction). 
Here $|\psi^{0}_{\bfk_\|}\rangle$ is single particle wavefunction of zero energy state for a given in-plane momentum, obtained by exactly diagonalizing the Bogoliubov Hamiltonian. 
We find that the wavefunction does decay more slowly that an exponential function inside the bulk, reflecting the effects of dipolar interaction when compared to the results of nearest neighboring pairing only. 
Besides, the decay slope is directly related to the energy gap  for a given in-plane momentum, $\bfk_\|$. 
For example, the energy gap is largest near $\bfk_\|= k_F$, and the  MF wavefunction decay with fast slope (shortest decay length). 
When $\bfk_\|$ is closed to $k_{\rm Max}$, this is, energy gap is closing, the single particle wavefunction of zero energy state becomes long tailed, as expected.

\section{Stability of MSZMs}
\label{stability}

In this section, we discuss the stability of the unpaired MSZMs obtained from the Kitaev ladder in the momentum space (see Eq.~(\ref{H_ladd1})). 
There are two aspects to study: 
first, if the system will be stable against the single particle perturbation on the edge surface; 
second, if the system is stable against the possible interaction between Majorana fermions due to the dipolar interaction. 
We will verify the first by applying the  time-reversal symmetry (TRS) on the current system, and argue the second from existing results in similar systems.

\subsection{Stability with respect to disorder or inhomogeneous potential}

In this paper, we concentrate on the situation when $\mu^\ast< 2t_z^\ast$ such that the 1D Kitaev ladders of ALL in-plane momentum are topological (see Fig.~\ref{fig:flatband long range}(a) and Eq.~(\ref{phase boundary2})). 
％
The most general single particle perturbation on these unpaired MFs in the bottom layer ($j=1$)  can be expressed as $H_{\rm dis}=H_1+H_2$, where
\begin{eqnarray}
H_1 &=& i\sum_{\bfk_\|\neq-\bfk'_\|}A(\bfk_\|,\bfk'_\|)\tilde{\gamma}^{(1)}_{\bfk_\|,1}\tilde{\gamma}^{(1)}_{\bfk_\|',1},
\label{H_1}
\notag\\
H_2 &=& i\sum_{\bfk_\|}A(\bfk_\|,-\bfk_\|)\tilde{\gamma}^{(1)}_{\bfk_\|,1}\tilde{\gamma}^{(1)}_{-\bfk_\|,1}.
\label{H_2}
\end{eqnarray}
Here $A(\bfk_\|,\bfk'_\|)$ is a general single particle potential due to local disorder or inhomogeneous potential $U_{\rm trap}(\bfr_\|)$ in the in-plane $(x-y)$ dimension. 
To avoid confusion, we also have defined all the unpaired MFs to be $\tilde{\gamma}^{(1)}_{\bfk_\|,1}$.

Note that these MF symbols are defined in the momentum space and therefore has a little bit different  properties compared to the regular MFs defined in real space. 
As described in Appendix~\ref{appA}, the MFs in momentum space have 
$\left(\tilde{\gamma}^{(1)}_{\bfk_\|,1}\right)^\dagger=\tilde{\gamma}^{(1)}_{-\bfk_\|,1}$, ${\cal T}\tilde{\gamma}^{(1)}_{\bfk_\|,1}{\cal T}^{-1}=\tilde{\gamma}^{(1)}_{-\bfk_\|,1}$. 
They satisfies the exchange properties: $\{\tilde{\gamma}^{(1)}_{\bfk_\|,1},\tilde{\gamma}^{(1)}_{\bfk_\|',1}\}=2\delta_{\bfk_\|,-\bfk_\|'}$. 
This is why we separate out terms with $\bfk'_\|=-\bfk_\|$ in above definition.

Starting from $H_1$ term, it is easy to find that the hermitian property of $H_1$ lead to the following constrain:
\be
A(\bfk_\|,\bfk'_\|)=A(-\bfk_\|,-\bfk'_\|)^\ast,
\label{A_hermitian}
\ee
because
\begin{eqnarray}
H_1=H_1^\dagger
&=&-i\sum_{\bfk_\|\neq-\bfk'_\|}A(\bfk_\|,\bfk'_\|)^\ast\tilde{\gamma}^{(1)}_{-\bfk_\|',1}\tilde{\gamma}^{(1)}_{-\bfk_\|,1}
\nonumber\\
&=&i\sum_{\bfk_\|\neq-\bfk'_\|}A(\bfk_\|,\bfk'_\|)^\ast\tilde{\gamma}^{(1)}_{-\bfk_\|,1}\tilde{\gamma}^{(1)}_{-\bfk_\|',1}
\nonumber\\
&=&i\sum_{\bfk_\|\neq-\bfk'_\|}A(-\bfk_\|,-\bfk'_\|)^\ast\tilde{\gamma}^{(1)}_{\bfk_\|,1}\tilde{\gamma}^{(1)}_{\bfk_\|',1}
\end{eqnarray}
%
On the other hand, the Time-Reversal Symmetry (TRS) of the system leads to
\begin{equation}
A(\bfk_\|,\bfk'_\|)=-A(-\bfk_\|,-\bfk'_\|)^\ast,
\label{A_TSR}
\end{equation}
because
\begin{eqnarray}
H_1&=&{\cal T}H_1{\cal T}^{-1}
\nonumber\\
&=& -i\sum_{\bfk_\|\neq-\bfk'_\|}A(\bfk_\|,\bfk'_\|)^\ast\tilde{\gamma}^{(1)}_{-\bfk_\|,1}\tilde{\gamma}^{(1)}_{-\bfk_\|',1}
\nonumber\\
&=&-i\sum_{\bfk_\|\neq-\bfk'_\|}A(-\bfk_\|,-\bfk'_\|)^\ast\tilde{\gamma}^{(1)}_{\bfk_\|,1}\tilde{\gamma}^{(1)}_{\bfk_\|',1}
\end{eqnarray}
One can see easily that the function $A(\bfk_\|,\bfk_\|')=0$ for all $\bfk'_\|\neq -\bfk_\|$, because Eq.~(\ref{A_hermitian}) and Eq.~(\ref{A_TSR}) cannot be satisfied simultaneously.

As for the remain term, $H_2$, the hermitian property requires that $A(\bfk_\|,-\bfk_\|)=-A(\bfk_\|,-\bfk_\|)^\ast$, because
\begin{eqnarray}
H_2=H_2^\dagger
&=&-i\sum_{\bfk_\|}A(\bfk_\|,-\bfk_\|)^\ast\tilde{\gamma}^{(1)}_{\bfk_\|,1}\tilde{\gamma}^{(1)}_{-\bfk_\|,1}
\end{eqnarray}
The TRS requires that $A(\bfk_\|,-\bfk_\|)=-A(-\bfk_\|,\bfk_\|)^\ast$, because
\begin{eqnarray}
H_2&=&{\cal T}H_2{\cal T}^{-1}
\nonumber\\
&=& -i\sum_{\bfk_\|}A(\bfk_\|,-\bfk_\|)^\ast\tilde{\gamma}^{(1)}_{-\bfk_\|,1}\tilde{\gamma}^{(1)}_{\bfk_\|,1}
\nonumber\\
&=& -i\sum_{\bfk_\|}A(-\bfk_\|,\bfk_\|)^\ast\tilde{\gamma}^{(1)}_{\bfk_\|,1}\tilde{\gamma}^{(1)}_{-\bfk_\|,1}
\end{eqnarray}
In other words, we obtain $A(\bfk_\|,-\bfk_\|)=A(-\bfk_\|,\bfk_\|)$. 
This implies that
\be
H_2 &=& i\sum_{\bfk_\|}A(\bfk_\|,-\bfk_\|)\tilde{\gamma}^{(1)}_{\bfk_\|,1}\tilde{\gamma}^{(1)}_{-\bfk_\|,1}
\nonumber\\
&=& i\sum_{\bfk_\|}A(-\bfk_\|,\bfk_\|)\tilde{\gamma}^{(1)}_{\bfk_\|,1}\tilde{\gamma}^{(1)}_{-\bfk_\|,1}
\nonumber\\
&=& i\sum_{\bfk_\|}A(\bfk_\|,-\bfk_\|)\tilde{\gamma}^{(1)}_{-\bfk_\|,1}\tilde{\gamma}^{(1)}_{\bfk_\|,1}
\nonumber\\
&=& \frac{i}{2}\sum_{\bfk_\|\neq0}A(\bfk_\|,-\bfk_\|)[\tilde{\gamma}^{(1)}_{\bfk_\|,1}\tilde{\gamma}^{(1)}_{-\bfk_\|,1}
\nonumber\\
&&+\tilde{\gamma}^{(1)}_{-\bfk_\|,1}\tilde{\gamma}^{(1)}_{\bfk_\|,1}]+iA(0,0)\tilde{\gamma}^{(1)}_{0,1}\tilde{\gamma}^{(1)}_{0,1}
\nonumber\\
&=& i\sum_{\bfk_\|\neq0}A(\bfk_\|,-\bfk_\|)+iA(0,0),
\ee
which is nothing but a constant only.

Therefore, we find that {\it any} single particle perturbation, $H_{\rm dis}=H_1+H_2$, are actually irrelevant to the energy or wavefunction Majorana fermions. 
This important consequence results from the fact that {\it all} unpaired MFs in the bottom layer, i.e. $\tilde{\gamma}^{(1)}_{\bfk,1}$, are transformed in the same way under time-reversal operation, while the other type of Majorana fermion, $\tilde{\gamma}^{(2)}_{\bfk,1}$, is not involved in this bottom layer at all. 
Therefore, the system composed by single type of MFs should be robust against any single particle perturbation when the time-reversal symmetry is considered, at least within the mean-field approximation we consider in this paper. 
This important result also applies to inhomogeneous trapping potential in the in-plane ($x-y$) direction, since it can be always written in the same form as $H_{\rm dis}$ even without periodic boundary condition.

Finally, we emphasize that the topological properties of a multi-layer structure discussed in this paper can be also understood as a 2D array of Kitaev chains {\it in real space}, where the inter-chain tunnelling makes the Majorana fermion mobile in the $(x-y)$ plane.
 In fact, Wakatsuki,  Ezawa,  and Nagaosa in Ref.~\cite{multichains} have shown that the Majorana zero modes of their multi-chain system can still exist and of multi-degeneracy when inter-chain tunnelling is turned on with an {\it open} boundary condition.
 In other words, the zero energy flat band does not depend on the assumption of periodic potential, completely consistent with results derived above. 
However, since our analytic results are confirmed by investigating the equivalent Kitaev two-leg ladders in the momentum space (see Fig.~\ref{fig:mutilayer}(b) and Eq.~(\ref{Hbcs})), our approach can be easily generalized to thermodynamic limit in a mesoscopic system.

\subsection{Stability with respect to the coupling between MSZMs and Bogoliubov quasi-particles}

The single particle perturbation in $H_{\rm dis}$ discussed above includes the coupling between MSZMs only.
 In principle,  the disorder potential may also couple MSZMs and Bogoliubov  quasi-particles in the bulk. 
From perturbation point of view, such coupling could be suppressed by the superfluid pairing gap, and the only possible nontrivial effects can appear only near the gapless ring: $|\bfk_\| |=k_{\rm Max}$.

However, we have to emphasize that the effects of such hybridization depends not only on the energy differences between particles, but also the occupation number in the relevant states.  
We will show that the occupation number of particles near the gapless ring at $|\bfk_\| |=k_{\rm Max}$ should be very small, and hence the coupling between MSZMs and the Bogoliubov quasi-particles should be also irrelevant to influence the topological properties. 

In order to see this, we first calculate the in-plane occupation number of particles near $|\bfk_\| |=k_{\rm Max}$, which is the Fermi momentum at $k_z=0$. 
Without loss of generality, the occupation number can be easily estimated by using periodic boundary condition. 
Within the standard BCS theory, the 3D occupation number in momentum space at zero temperature is given by, $n^{3D}_{\bfk}=\frac{1}{2}\left[1-\xi_\bfk/E_\bfk^{(+)}\right]$, where $E_\bfk^{(+)}$ is Bogoliubov excitation energy in Eq.~(\ref{excitation}). 
Therefore  the 2D occupation number of particles for a given in-plane momentum is given by 
\be
n_{\bfk_\|}^{2D}&\equiv &\int_{-\pi/d}^{\pi/d}\frac{dk_z}{2\pi}n^{3D}_{\bfk} 
= \frac{1}{2} \int_{-\pi/d}^{\pi/d}\frac{dk_z}{2\pi}\left[1-\frac{\xi_\bfk}{E_\bfk^{(+)} }\right]
\nonumber\\
&\sim &\int_{-\pi/d}^{\pi/d}\frac{dk_z}{2\pi}\frac{1}{2}\left[1-\frac{\xi_\bfk}{|\xi_\bfk|}\right]
\nonumber\\
&=&\int_{-k_0}^{k_0}\frac{dk_z}{2\pi}=\frac{1}{\pi d}\cos^{-1}\left(\frac{\bfk^2_\|/2m^\ast-\mu^\ast}{2t_z^\ast}\right),
\ee
where $k_0\equiv \frac{1}{d}\cos^{-1}\left((\bfk^2/2m^\ast-\mu^\ast)/2t_z^\ast\right)$. From the second line, we have assumed the gap is small and negligible compared to the chemical potential, so that certain analytic estimate can be derived. 

It is easy to see that $n^{2D}_{\bfk_\|}\to \frac{1}{\pi d}\cos^{-1}(1)=0$, when $ |\bfk_\| |\to k_{\rm Max}=\sqrt{2m^\ast(\mu^\ast+2t_z^\ast)}$. 
In other words, we could reasonably expect that there is almost {\it no} Bogoliubov particles or Majorana fermions occupied near the gapless ring at $|\bfk_\| |=k_{\rm Max}$. 
Therefore, after considering the negligible occupation number of particles near the gapless ring we believe it is reasonable to neglect  the coupling  between MSZMs and Bogoliubov quasi-particles. The topological system we proposed should be still stable in the thermal dynamic limit.

Finally, we note that when $\mu^\ast > 2t_z^\ast$, an additional gapless ring appears at $|\bfk_\| |=k_{\rm Min}$ ,where the occupation number there is still  finite and like the case near $|\bfk_\| | = k_{\rm Max}$.
As a result, certain coupling between the Majorana surface modes and the Bogoliubov quasi-particles may be possible. 
The topological property for the later case may still exist, but should be further investigated and therefore not the situation we will consider in this paper.

\begin{figure}[htb!]
\centering
\includegraphics[width=0.5\textwidth]{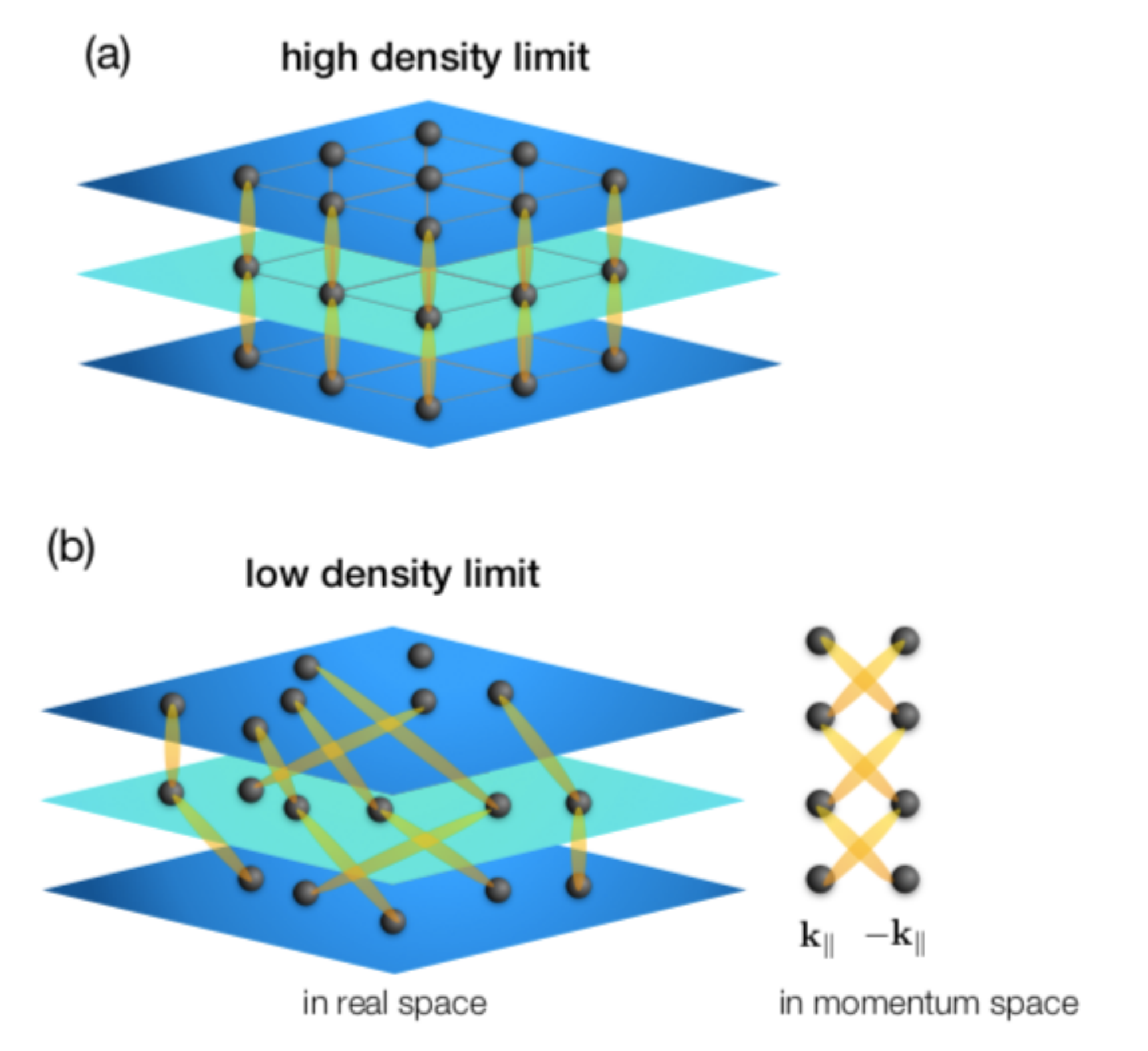}
\caption{ 
Schematic plots of polar molecules in a multilayer system in  high density limit (a) and in  low density limit (b) respectively. 
In high density limit, strong repulsive dipolar interaction makes the molecules to form a triangular crystal in each layer. 
On the other hand, in low density limit, molecules are free-moving without fixed relative position with each other.
Therefore when inter-layer attraction is increased by reducing inter-layer distance, the former can form a 2D array of Kitaev chains with a vertical pairing between each other in different layers, while the latter will form cross pairing between in-plane momenta $\bfk_{\|}$ and $-\bfk_{\|}$ as in a Kitaev ladder (see also Fig.~\ref{fig:mutilayer}(b), Eqs.~(\ref{Hbcs}) and (\ref{H_ladd1}) ). 
Here for simplicity, we just show the pairing between molecules in the nearest neighboring layers (denoted by yellow elliptical curves).
}
\label{fig:density}
\end{figure}

\subsection{Stability with respect to in-plane repulsive interaction}

In this section, we briefly address the interaction effects between MSZMs. 
We start from considering results in both high density and low density regimes, where the pairing is better-defined in the real space for the former, while better-defined in the momentum space for the later. 
We will argue that they should exhibit the same topological phase even in the intermediate regime at least within the meanfield approximation. 
Full analytic or numerical calculation beyond the meanfield level is beyond the scope of this paper and should be studied more carefully in the future.

Our argument starts from the high density limit (not addressed in this paper), where the averaged inter-particle distance between polar molecules in the same layer is shorter than the inter-layer spacing.
Hence the horizontal (in-plane) kinetic energy is suppressed by the strong repulsive dipolar interaction.
 As a result, polar molecules in each layer can be self-assembled to be a triangular lattice, and the relative position of these lattice points (molecules) are mostly frozen and aligned to each other (See. Fig.~\ref{fig:density}(a) and Ref.~\cite{FabioDW, YouDW}). 
 When a small amount of vacancies appears for certain chemical potential, $p$-wave superfluidity appears within each 1D chain along the normal ($z$) direction, and the system becomes almost equivalent to a planar array of 1D Kitaev chains. 
 The topological properties therefore should be equivalent to a 2D array of Kitaev chain with interlayer pairing, i.e.,  the BDI class (as the regular Kitaev chain) with localized Majorana fermions at the edge forming a flat band as studied in Ref. \cite{multichains}.

In the dilute limit (which is what we studied in this paper), the intra-layer dipolar repulsion is actually much weaker compared to the in-plane kinetic energy.
It is well-justified to apply Landau-Fermi liquid theory in each 2D layer before the inter-layer pairing is introduced (see Refs. \cite{Pikovski2010, DasSarma}). 
Such a weak intra-layer repulsion just renormalizes the effective mass and chemical potential, and the topological properties of such a multi-layer system can be understood as a bundle of 1D Kitaev ladder in the momentum space, as we addressed in the previous part of this paper. 
 The resulting topological phase is also belong to BDI class with quantum degenerate Majorana surface modes in the top and bottom layer if Eq.~(\ref{phase boundary2} is satisfied, see Fig.~\ref{fig:density}(b).

In the middle range of the in-plane density, however, intra-layer repulsion is comparable to the in-plane kinetic energy and hence the topological properties have to be identified beyond single particle (meanfield) picture. 
A relevant study on the interaction effects of Majorana fermions has been discussed~\cite{interaction1,interaction2}.
For example, Fidkowski and Kitaev have shown that  the BDI class with $\mathbb{Z}$ topological invariance shall be broken to $\mathbb{Z}_8$ when a short ranged four-point interaction is included~\cite{interaction1,interaction2} .

But we note that in our present system the long-ranged nature of dipolar interaction also makes the system quite different from the short-ranged case studied by Fidkowski and Kitaev.  
More specifically, we could write down the most general mutual interaction term between four fermions in the bottom layers to be
\be
H_\gamma&=&\frac{1}{\Omega}\sum_{\bfk_{\|,i}}{}'U_\gamma(\{\bfk_{\|,i}\}) \tilde{\gamma}^{(1)}_{\bfk_{\|,1},1}\tilde{\gamma}^{(1)}_{\bfk_{\|,2},1}\tilde{\gamma}^{(1)}_{\bfk_{\|,3},1}\tilde{\gamma}^{(1)}_{\bfk_{\|,4},1},
\ee
where $U_\gamma(\bfk_{\|,1},\cdots,\bfk_{\|,4})$ is some complicated interaction matrix element between MFs, $\tilde{\gamma}^{(1)}_{\bfk_{\|,i},1}$. $\sum_{\bfk_{\|,i}}'$ indicates a summation of $\bfk_{\|,i}$ with the conservation of total momentum, $\bfk_{\|,1}+\cdots+\bfk_{\|,4}=0$.

In order to recover Fidkowski and Kitaev's result for $Z_8$ symmetry, $U_\gamma$ has to be a constant, independent of the in-plane momenta. 
This indicates that the bare interaction between polar molecules has to be a short-ranged interaction in the real space, which cannot be realistic/effective for our spinless fermionic polar molecules.
In fact, fermionic SPT phases are not totally understood yet in 3D system, although it is argued that there exists a bulk SPT and surface symmetry enriched TO anomaly matching in a 3D bosonic weak SPT phase \cite{Bosonic_SPT}.
It is a very interesting and challenging subject for the future study. The multi-layer structure as well as the Majorana surface modes proposed in this paper suggest a very promising system to investigate the many-body physics of these non-Abelian particles.

Finally, we note that in Ref.~\cite{multichains} the authors also studied the case when the inter-chain pairing is included. 
Within the meanfield (BCS) approximation, they find that the multi-degenerate Majorana zero modes are still present and  stable if the inter-chain pairing is of the same phase as the intra-chain pairing. 
These degenerate zero modes becomes unstable (and hence with a finite dispersion) {\it only} when the the time reversal symmetry is broken ( i.e. the inter-chain pairing is of different phase as the intra-chain pairing). 
Therefore, it is reasonable to expect that the quantum degenerate Majorana surface zero modes (in BDI class) observed in our multi-layer system may be still stable against the in-plane repulsive interaction between polar molecules, where the time-reversal symmetry is still strictly preserved.

\begin{figure}[htb!]
\centering
\includegraphics[width=0.5\textwidth]{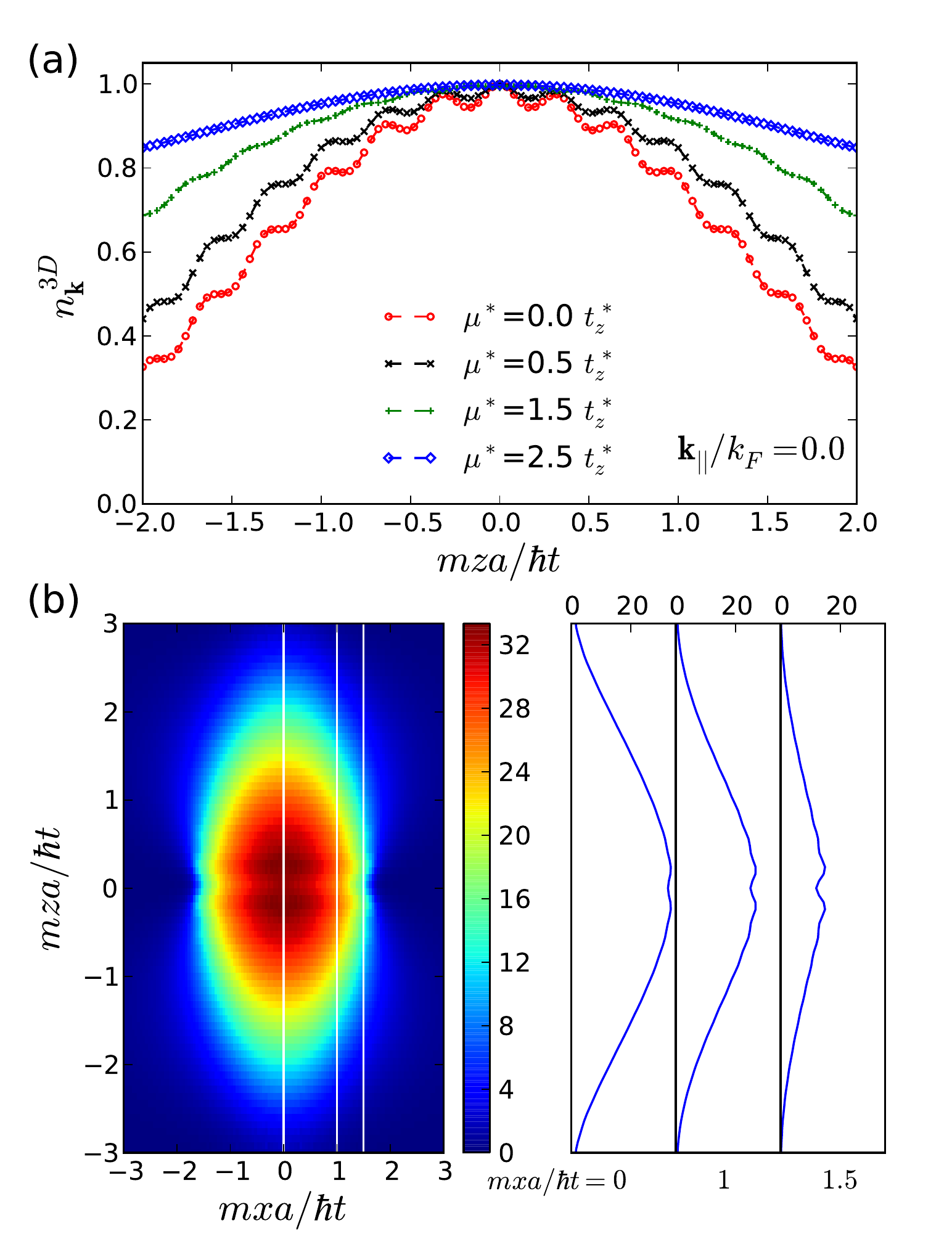}
\caption{
(a) The momentum distribution in 3D space $n^{3D}_{\bfk}$ as a function of the final position $z$ with  fixed $\bfk_\|/k_F = 0$ for various $\mu^\ast$. 
(b) Integrated momentum distribution along $y$ direction, $n_{\rm ToF}(x,z)$, for time-of-flight measurement, showing a possible feature on the interference pattern in the time-of-flight experiment.
Here we consider $\mu^\ast= \Delta_{\bfk_\|}^{(1)}=t_z^\ast$. The three cross sections $({mx}/{\hbar t} = 0,1,1.5)$ at different horizontal direction are shown in the right. 
}
\label{fig:n_k}
\end{figure}

\section{Experimental measurement} 
\label{experiment}

There have been several proposals to measure MFs. 
The simplest way is by observing the long-distance correlation between MFs in the two ending layers (for example, see \cite{Zoller_1Dladder}), which should cause a short interference period (i.e.  $\propto\cos(k_z Ld)$) in the momentum space (i.e. in the long time approximation of the time-of-flight experiments).
Here $Ld$ is the distance between the top and bottom layers. 
We emphasize that such measurement for 1D Kitaev chain model suffers a serious noise-to-signal ratio \cite{Zoller_1Dladder}, since only  ``one ``  Majorana mode exists at each end of the chain.
Besides, the fact that no long-ranged order can exist in 1D system also makes it difficult to expect a significant signal. 
In our multi-layer system, however, unpaired MFs are macroscopically degenerate near the surface layers, and therefore the experimental measurement of these MSZMs becomes much more promising. 

The unpaired MFs  can be characterized by the non-local fermionic correlations which can be detected in  time-of-flight (TOF) imaging, as illustrated in Fig.~\ref{fig:n_k}.
 We then study the time-of-flight image for a free expansion time $t$, obtained by integrating over the $y$ direction, i.e. 
\be
n_{\rm ToF}(x,z)&\propto &\left|\tilde{\omega}_0\left(\frac{mz}{\hbar t}\right)\right|^2 \int_{-\infty}^\infty dy \; n^{3D}_{m\bfr/\hbar t}
\ee
where the final position $\bfr=\hbar\bfk t/m$, and $\tilde{\omega}_0(k_z)$ is the fourier transform of the lowest band Wannier function. 
$n^{3D}_{\bfk}$ is the momentum distribution in 3D space,
\be
n^{3D}_{\bfk}=n^{3D}_{\bfk_\|,k_z}\equiv 
\sum_{j,j'=1}^L\langle G|c^\dagger_{\bfk_\|,j}c^{}_{\bfk_\|,j'}|G\rangle\,e^{i(j-j')dk_z},
\ee
where $|G\rangle$ is the ground state wavefunction obtained by diagonalizing the meanfield Hamiltonian in Eq.~(\ref{Hbcs}) with a long-ranged pairing. 

We first calculate the momentum distribution $n^{3D}_{\bfk}$ with fixed in-plane momentum $\bfk_\|=0$ with various the chemical potentials. 
By ramping the chemical potential, we can detect the transition point, since the oscillations disappear
 if we approach the transition point at $\mu^\ast = 2 t_z^\ast $ (see  Fig.~\ref{fig:n_k} (a) ) with $\bfk_\| /k_F=0$.
This numerical result agrees with Eq.~(\ref{phase boundary2}). 
In the Fig.~\ref{fig:n_k} (b), we show the TOF signals for the case with $\mu^\ast= \Delta_{\bfk_\|}^{(1)}=t_z^\ast$, proving that the oscillations will still be present  in the topological regime. 
As we can see that, when the system is in the topological regime, additional interference fringes emerge due to the long-ranged correlation between macroscopically degenerate Majorana surface modes in the top and bottom layers. 
The contrast of such interference to its average density is larger when away from the center, making it possible to be measured in the present experiments.

\section{Conclusion}
\label{conclusion}

In this paper, we propose that a novel topological state can be prepared and observed in a stack of 2D layers with fermionic polar molecules polarized to the normal direction. 
The associated quantum degenerate Majorana surface modes appear in the surface layers due to the 3D $p$-wave superfluidity, confirmed by the winding number, entanglement spectrum and entanglement entropy numerics. 
We also show the parameter regime to find MFs in the current experiments. Our work paves the way for future investigation of the many-body physics with non-Abelian statistics under time-reversal symmetry.

\section*{Acknowledgment}
We thank C.-Y Mou, S.-J. Huang, Y.-P. Huang, J-S. You and G. Juzeliunas for valuable discussion.

\appendix
\wbe
\section{Exact solution for a special case of Kitaev ladder in momentum space} 
\label{appA}

In this section, we will derive the exact solution for a special case of Kitaev ladder in momentum space, i.e. $\mu^\ast-\bfk_\|^2/2m^\ast=0$ and $\Delta_{\bfk_\|}^{(1)}=t_z^\ast$ in Eq.~(\ref{Hbcs}). 
All longer-ranged pairing is assumed to be zero for simplicity.

\subsection{Majorana fermions in momentum space }

To see how MFs appears in $H_{\bfk_\|}^{\rm ladd}$, we first note that the definition of Majorana fermions (MFs) in real space is different from that in momentum space. 
For a general fermionic field operator $C(x)$ at the position $x$, its relation with MF operator, $\Gamma(x)$, are following:
\begin{equation}
C(x) = \frac{1}{2} \left[ \Gamma_1(x)+ i \Gamma_2(x) \right], \ \ \
C^{\dagger} (x) = \frac{1}{2} \left[ \Gamma_1(x)- i \Gamma_2(x) \right], 
\end{equation}
and hence 
\begin{equation}
\Gamma_1(x) = C(x) + C^{\dagger} (x), \ \ \    
\Gamma_2(x) = -i \left[ C(x) - C^{\dagger} (x) \right].  
\end{equation}

Therefore, by doing Fourier Transformation, the Majorana operators in momentum space can be given 
\begin{eqnarray}
 \Gamma_{1,k} & = & \int \Gamma_1(x) e^{-i kx} dx  = \int  C(x) e^{-i kx} dx   + \int  C^{\dagger} (x) e^{-i kx} dx   
                        = C_{k} + C^{\dagger}_{-k},
\\                      
 \Gamma_{2,k}  &=&  -i \left[ C_k -  C^{\dagger} _{-k}  \right].                           
\end{eqnarray}
We note that now the MF operator $\Gamma_{1/2,k}$ are composed by Dirac fermionic operator at momentum $k$ and its anti-particles in the momentum $-k$.   

From above definition, the hermitian and time-reversal operation of the Majorana fermions in momentum space become:
\be
\Gamma_{1,k}^\dagger &=& C_{k}^\dagger + C^{}_{-k}=\Gamma_{1,-k}
\\
\Gamma_{2,k}^\dagger &=& i\left[C_{k}^\dagger - C^{}_{-k}\right]=\Gamma_{2,-k}
\\
{\cal T}\,\Gamma_{1,k}{\cal T}^{-1} &=& C_{-k}^\dagger + C^{}_{k}=\Gamma_{1,-k}
\\
{\cal T}\,\Gamma_{2,k}{\cal T}^{-1} &=&  i\left[C_{-k} - C^{\dagger}_{k}\right]=-\Gamma_{2,-k},
\ee
where ${\cal T}$ is the time reversal operator. We have used the fact that ${\cal T}C(x){\cal T}^{-1}=C(x)$ and ${\cal T}C_k{\cal T}^{-1}=C_{-k}$ to preserve the time reversal symmetry.

\subsection{ The Hamiltonian in Majorana fermion basis}

Using similar notation, we can define MFs in our present case as following:
\begin{eqnarray}
c_{{\bfk_\|},j} = \frac{1}{2} \Big[   \gamma_{{\bfk_\|},j}^{(1)}  + i \;   \gamma_{{\bfk_\|},j}^{(2)}       \Big],    
c^{\dagger}_{{\bfk_\|},j} = \frac{1}{2} \Big[   \gamma_{-{\bfk_\|},j}^{(1)}- i \;  \gamma_{-{\bfk_\|},j}^{(2)}        \Big],   
c_{-{\bfk_\|},j}  =  \frac{1}{2} \Big[   \gamma_{-{\bfk_\|},j}^{(1)}  + i \;   \gamma_{-{\bfk_\|},j}^{(2)}        \Big],    
c^{\dagger}_{-{\bfk_\|},j} = \frac{1}{2} \Big[   \gamma_{{\bfk_\|},j}^{(1)} - i \;  \gamma_{{\bfk_\|},j}^{(2)}    \Big]     
\end{eqnarray}
 
Since we are considering a special case: $\Delta_{\bfk_\|}=t_z^\ast$ and $\bfk_\|^2/2m^\ast=\mu^\ast$,  Hamiltonian in Eq.~(\ref{Hbcs}) can then be written as
(here we consider nearest pairing $\Delta_{\bfk_\|}\equiv\Delta_{\bfk_\|}^{(1)}$ only and assume longer-ranged pairing, $\Delta_{\bfk_\|}^{(j)}=0$ for $j>1$): 
\begin{eqnarray}
H_{\bfk_\|}^{\rm ladd} &=&  -t_z  \sum_{j=1}^{L-1}  
\left[   c^{\dagger}_{{\bfk_\|},j} c_{{\bfk_\|},j+1} +c^{\dagger}_{-{\bfk_\|},j} c_{-{\bfk_\|},j+1} + 
        c^{\dagger}_{{\bfk_\|},j} c^{\dagger}_{-{\bfk_\|},j+1} +  
       c^{\dagger}_{-{\bfk_\|},j} c^{\dagger}_{{\bfk_\|},j+1} +    
       \right.
\nonumber\\
&&\left.+c^{\dagger}_{{\bfk_\|},j+1} c_{{\bfk_\|},j} 
      +c^{\dagger}_{-{\bfk_\|},j+1} c_{-{\bfk_\|},j}    +
        c_{-{\bfk_\|},j+1} c_{{\bfk_\|},j} +c_{{\bfk_\|},j+1} c_{-{\bfk_\|},j},                                               
             \right]  
\end{eqnarray}
where $L$ is the number site/layers along the $z$-direction. 

After some straightforward algebra, the first four terms in $\big[  \quad  \big]$ of Hamiltonian is given by 
\begin{eqnarray}
&&\frac{1}{4} ( \gamma^{(1)}_{-{\bfk_\|},j} - i \;  \gamma^{(2)}_{-{\bfk_\|},j} ) (  \gamma^{(1)}_{{\bfk_\|},j+1}  + i \;   \gamma^{(2)}_{{\bfk_\|},j+1}  )+            
 \frac{1}{4} ( \gamma^{(1)}_{{\bfk_\|},j} - i \;  \gamma^{(2)}_{{\bfk_\|},j} )(   \gamma^{(1)}_{-{\bfk_\|},j+1}  + i \;   \gamma^{(2)}_{-{\bfk_\|},j+1} )
\nonumber\\
&&+ \frac{1}{4} ( \gamma^{(1)}_{-{\bfk_\|},j} - i \;  \gamma^{(2)}_{-{\bfk_\|},j} ) ( \gamma^{(1)}_{{\bfk_\|},j+1} - i \;  \gamma^{(2)}_{{\bfk_\|},j+1} )+                   
 \frac{1}{4} ( \gamma^{(1)}_{{\bfk_\|},j} - i \;  \gamma^{(2)}_{{\bfk_\|},j} )( \gamma^{(1)}_{-{\bfk_\|},j+1} - i \;  \gamma^{(2)}_{-{\bfk_\|},j+1} )      
\nonumber\\
&=&\frac{1}{4} (  \gamma^{(1)}_{-{\bfk_\|},j} \gamma^{(1)}_{{\bfk_\|},j+1} + \gamma^{(2)}_{-{\bfk_\|},j}   \gamma^{(2)}_{{\bfk_\|},j+1}  )  +
    \frac{i}{4} (  \gamma^{(1)}_{-{\bfk_\|},j}  \gamma^{(2)}_{{\bfk_\|},j+1}  -   \gamma^{(2)}_{-{\bfk_\|},j} \gamma^{(1)}_{{\bfk_\|},j+1}  )  
\nonumber\\
&&+ \frac{1}{4} (\gamma^{(1)}_{{\bfk_\|},j} \gamma^{(1)}_{-{\bfk_\|},j+1} +  \gamma^{(2)}_{{\bfk_\|},j} \gamma^{(2)}_{-{\bfk_\|},j+1} )  +
    \frac{i}{4} (  \gamma^{(1)}_{{\bfk_\|},j} \gamma^{(2)}_{-{\bfk_\|},j+1} -   \gamma^{(2)}_{{\bfk_\|},j} \gamma^{(1)}_{-{\bfk_\|},j+1} )  
\nonumber\\
&&+\frac{1}{4} (\gamma^{(1)}_{-{\bfk_\|},j} \gamma^{(1)}_{{\bfk_\|},j+1} -  \gamma^{(2)}_{-{\bfk_\|},j} \gamma^{(2)}_{{\bfk_\|},j+1} ) +
     \frac{-i}{4} (\gamma^{(1)}_{-{\bfk_\|},j}\gamma^{(2)}_{{\bfk_\|},j+1} +  \gamma^{(2)}_{-{\bfk_\|},j}\gamma^{(1)}_{{\bfk_\|},j+1} ) 
\nonumber\\
&&+ \frac{1}{4} (\gamma^{(1)}_{{\bfk_\|},j}  \gamma^{(1)}_{-{\bfk_\|},j+1}  -  \gamma^{(2)}_{{\bfk_\|},j} \gamma^{(2)}_{-{\bfk_\|},j+1}  )+
      \frac{-i}{4} ( \gamma^{(1)}_{{\bfk_\|},j} \gamma^{(2)}_{-{\bfk_\|},j+1} +  \gamma^{(2)}_{{\bfk_\|},j} \gamma^{(1)}_{-{\bfk_\|},j+1}) 
\end{eqnarray}

Similarly, the last four terms in $\big[  \quad  \big]$ is given by 
\begin{eqnarray}
&& \frac{1}{4} ( \gamma^{(1)}_{-{\bfk_\|},j+1} - i \;  \gamma^{(2)}_{-{\bfk_\|},j+1} ) (  \gamma^{(1)}_{{\bfk_\|},j}  + i \;   \gamma^{(2)}_{{\bfk_\|},j}  ) +          
 \frac{1}{4} ( \gamma^{(1)}_{{\bfk_\|},j+1}  - i \;  \gamma^{(2)}_{{\bfk_\|},j+1}  )(   \gamma^{(1)}_{-{\bfk_\|},j}  + i \;   \gamma^{(2)}_{-{\bfk_\|},j}  ) 
\nonumber\\
&&+\frac{1}{4} ( \gamma^{(1)}_{-{\bfk_\|},j+1}  + i \;  \gamma^{(2)}_{-{\bfk_\|},j+1} ) ( \gamma^{(1)}_{{\bfk_\|},j}  + i \;  \gamma^{(2)}_{{\bfk_\|},j}  )+                   
\frac{1}{4} ( \gamma^{(1)}_{{\bfk_\|},j+1}  +  i \;  \gamma^{(2)}_{{\bfk_\|},j+1}  )( \gamma^{(1)}_{-{\bfk_\|},j} + i \;  \gamma^{(2)}_{-{\bfk_\|},j} )  
\nonumber\\
&=& \frac{1}{4} (  \gamma^{(1)}_{-{\bfk_\|},j+1}   \gamma^{(1)}_{{\bfk_\|},j}  + \gamma^{(2)}_{-{\bfk_\|},j+1}    \gamma^{(2)}_{{\bfk_\|},j}   )  +
    \frac{i}{4} (  \gamma^{(1)}_{-{\bfk_\|},j+1} \gamma^{(2)}_{{\bfk_\|},j}   -   \gamma^{(2)}_{-{\bfk_\|},j+1}  \gamma^{(1)}_{{\bfk_\|},j}   )  
\nonumber\\
&&+\frac{1}{4} (\gamma^{(1)}_{{\bfk_\|},j+1}  \gamma^{(1)}_{-{\bfk_\|},j}  +  \gamma^{(2)}_{{\bfk_\|},j+1}   \gamma^{(2)}_{-{\bfk_\|},j}   )  +
    \frac{i}{4} (  \gamma^{(1)}_{{\bfk_\|},j+1}  \gamma^{(2)}_{-{\bfk_\|},j}  -   \gamma^{(2)}_{{\bfk_\|},j+1}   \gamma^{(1)}_{-{\bfk_\|},j}   )  
\nonumber\\
&&+ \frac{1}{4} (\gamma^{(1)}_{-{\bfk_\|},j+1}   \gamma^{(1)}_{{\bfk_\|},j}  -  \gamma^{(2)}_{-{\bfk_\|},j+1}   \gamma^{(2)}_{{\bfk_\|},j}   ) +
     \frac{i}{4} (\gamma^{(1)}_{-{\bfk_\|},j+1} \gamma^{(2)}_{{\bfk_\|},j}  +  \gamma^{(2)}_{-{\bfk_\|},j+1} \gamma^{(1)}_{{\bfk_\|},j}  ) 
\nonumber\\
&&+\frac{1}{4} (\gamma^{(1)}_{{\bfk_\|},j+1}  \gamma^{(1)}_{-{\bfk_\|},j}   -  \gamma^{(2)}_{{\bfk_\|},j+1}  \gamma^{(2)}_{-{\bfk_\|},j}    )+
      \frac{i}{4} ( \gamma^{(1)}_{{\bfk_\|},j+1}  \gamma^{(2)}_{-{\bfk_\|},j}  +  \gamma^{(2)}_{{\bfk_\|},j+1}  \gamma^{(1)}_{-{\bfk_\|},j}  )   
\end{eqnarray}

Now using the the anti-commutation relations,{$\{\gamma^{(a)}_{j,\bfk_\|},\gamma^{(b)}_{j',\bfk_\|'}\}=2\delta_{j,j'}\delta_{a,b}\delta_{\bfk_\|,-\bfk_\|'}$} ($a,b=1,2$), 
we can obtain the Hamiltonian with special case $\Delta_{\bfk_\|}=t_z$, 
\begin{equation}
H_{\bfk_\|}^{\rm ladd}=it_z\sum_{j=1}^{L-1}\left(\gamma_{\bfk_\|,j}^{(2)}\gamma_{-\bfk_\|,j+1}^{(1)}+\gamma_{-\bfk_\|,j}^{(2)}\gamma_{\bfk_\|,j+1}^{(1)}\right).
\end{equation}

\section{ BDI class and Winding number}  \label{appB}
In order to specify the TO, we have to check the symmetry first. 
\begin{equation}
 {\cal T}H(\bfk) {\cal T}^{-1}=H(-\bfk),
 {\cal C}H(\bfk) {\cal C}^{-1}=-H(-\bfk),
{\cal S}H(\bfk) {\cal S}^{-1}=-H(\bfk).
\label{symmetry}
\end{equation}
Where  ${\cal T}$, ${\cal C}$, and  ${\cal S}$ are the anti-unitary time reversal symmetry (complex conjugate for spinless model), particle-hole symmetry and chiral symmetry (product of ${\cal T}$ and ${\cal C}$).  
It belongs to BDI class and topological phases is characterized by the $\mathbb{Z}$ index. This topological state is protected by chiral symmetry.

The BdG Hamiltonian can be unitary transformed to off-diagonal form,  $H_{BdG}=\frac{i}{4}\sum_{\bfk}   \Gamma^{\dagger}_{\bfk}  \left[ 
  \begin{array}{cc}   
    0 & v_{\bfk} \cr      
    v_{\bfk}^{\dagger} & 0 \cr  
  \end{array}
\right]    \Gamma_{\bfk},$  where $\mu_{\bfk_\|}^\ast=\mu^\ast+\frac{k_{\parallel}^2}{2m^\ast}$, $v_{\bfk}=-\mu_{\bfk_\|}^\ast-2t^\ast_z\cos (k_z d)-2i\Delta\sin (k_z d)$ and $v_\bfk=R(\bfk)\mathrm{e}^{i\theta (\bfk)}$.
Then, the corresponding Q-matrix can be given,
$Q_{\bfk} =\left[                 
  \begin{array}{cc}   
    0 & q_{\bfk}  \cr      
    q_{\bfk} ^{\dagger} & 0 \cr  
  \end{array}
\right], $ where $q_{\bfk} =\frac{v_{\bfk} }{|v_{\bfk} |}=e^{i \theta_{\bfk}}$. 
The relevant space is the U(N) unitary group. $q_{\bfk}$ is a mapping from z-direction BZ to U(N), which topologically classified by first homotopy group, $\Pi_1[U(N)]=\mathbb{Z}$, characterized by the winding number $W$, see Ref.~\cite{classification}.
The winding number can be calculated as following: 
\begin{eqnarray}
W_{\bfk_\|} &\equiv & \int_{-\pi/d}^{\pi/d}\frac{dk_z}{2\pi}\partial_{k_z}\theta_{\bfk_\|,k_z} 
= \int_{-\pi/d}^{\pi/d}\frac{dk_z}{2\pi}\frac{\partial_{k_z}[-\cos\theta_{\bfk_\|,k_z}]}{\sin\theta_{\bfk_\|,k_z}}  
\nonumber\\
&=& \int_{-\pi/d}^{\pi/d}\frac{dk_z}{-2\pi}\frac{\partial_{k_z}[-(\mu_{\bfk_\|}+2t_z \cos (k_z d))/R(\bfk))]}{-[2\Delta_{\bfk_\|} \sin (k_z d)]}/R(\bfk)    
\nonumber\\
&=& \frac{1}{4\pi}\int_{-\pi}^{\pi}dk_z\frac{1}{\Delta_{\bfk_\|}\sin k_z }   \;  \left[ 2t_z \sin k_z + 2\frac{[\mu_{\bfk_\|}+2t_z \cos k_z]}{R(\bfk)^2} 
[2\Delta_{\bfk_\|}^2\sin k_z\cos k_z-t_z\sin k_z (\mu_{\bfk_\|}+2t_z \cos k_z)]\right]  
\nonumber\\
&=& \frac{1}{2\pi\Delta_{\bfk_\|}}\int_{-\pi}^{\pi}dk_z\frac{4\Delta_{\bfk_\|}^2 t_z+2\Delta_{\bfk_\|}^2\mu_{\bfk_\|}\cos k_z}{[\mu_{\bfk_\|}+2t_z \cos k_z]^2+4\Delta_{\bfk_\|}^2\sin^2 k_z }
\end{eqnarray}

Changing the variables $ z=e^{i k_z d}$ and let $J_{1}=\frac{t_z+\Delta_{\bfk_\|}}{\mu_{\bfk_\|}}$, $J_{2}=\frac{t_z-\Delta_{\bfk_\|}}{\mu_{\bfk_\|}}$, and $J_{\pm}=J_{1}\pm J_{2}$,
\begin{eqnarray}
W_{\bfk_\|} &=& \frac{1}{2\pi i}\oint \frac{dz}{z} \Delta_{\bfk_\|}\frac{4 t_z+\mu_{\bfk_\|}(z+\frac{1}{z} )}{[\mu_{\bfk_\|}+t_z (z+\frac{1}{z} )]^2-\Delta_{\bfk_\|}^2(z-\frac{1}{z} )^2} 
\nonumber\\
&=&\frac{\Delta_{\bfk_\|}}{2\pi i}\oint dz \frac{\mu_{\bfk_\|} z^2+4 t_z z+ \mu_{\bfk_\|}}{[(t_z+\Delta_{\bfk_\|})z^2+\mu_{\bfk_\|} z+ (t_z-\Delta_{\bfk_\|})][(t_z-\Delta_{\bfk_\|})z^2+\mu_{\bfk_\|} z+ (t_z+\Delta_{\bfk_\|})]} 
\nonumber\\
&=&\frac{J_{-}}{2}\oint \frac{dz}{2\pi i} \frac{ z^2+2 J_{+} z+ 1}{[J_{1}z^2+ z+ J_{2}][J_{2}z^2+ z+ J_{1}]} 
\nonumber\\
&=& \frac{J_{-}}{2}\oint \frac{dz}{2\pi i} \frac{ z^2+2 J_{+} z+ 1}{J_{1}J_{2}[(z\hspace{-0.2em}-\hspace{-0.2em}Z_1)(z\hspace{-0.2em}-\hspace{-0.2em}Z_2)(z\hspace{-0.2em}-\hspace{-0.2em}Z_3)(z\hspace{-0.2em}-\hspace{-0.2em}Z_4)]}
\label{winding number calculation}
\end{eqnarray}
Four poles of the function are $Z_{1,2}=\frac{-1 \pm \sqrt{1-4 J_1 J_2} }{2 J_1}$ and $Z_{3,4}=\frac{-1 \pm \sqrt{1-4 J_1 J_2} }{2 J_2}$ respectively.

It can be shown that  $Z_{3,4}=1/Z_{1,2}$.
By Cauchy's residue theorem, the winding number and the phase boundary can be determined. 
There are three different situations:
\begin{eqnarray}
A: &&\mathrm{If} \hspace{0.5em} |Z_{1}|<1\hspace{0.5em}  \mathrm{and} \hspace{0.5em} |Z_{2}|<1, \hspace{0.5em}   
            \qquad \Longrightarrow   \hspace{0.5em}W=1. 
\\
B: && \mathrm{If} \hspace{0.5em} |Z_{3}|<1\hspace{0.5em}  \mathrm{and} \hspace{0.5em} |Z_{4}|<1, \hspace{0.5em}   
            \qquad \Longrightarrow  \hspace{0.5em}W=-1.  
\\
C: && \mathrm{If} \hspace{0.5em} |Z_{1}|<1\hspace{0.5em}  \mathrm{and} \hspace{0.5em} |Z_{3}|<1,  \hspace{0.5em} \mathrm{or} \hspace{0.5em} |Z_{2}|<1 \hspace{0.5em}\mathrm{and} \hspace{0.5em} |Z_{4}|<1 ,\hspace{0.5em}   
 \qquad \Longrightarrow  \hspace{0.5em}W=0. 
\label{condition for TO}
\end{eqnarray}
They are three different gapped phases, where conditions $A$ and $B$ are corresponding to two different non-trivial topological phases.  
The corresponding $Q$ matrix element are winding around origin of complex plane counterclockwise and clockwise respectively. 
Phase boundaries are determined by $ |Z_{1,2}|= |Z_{3,4}|=1$, and hence  $\mu^\ast_{\bfk_\|}=\pm 2t_{z}$, which is consistent with condition for gap closing. 
Therefore, we can solve the topological non-trivial conditions for $A$ and $B$ situations, and obtain
\begin{equation}
|\mu^\ast_{\bfk_\|}|=|-\mu^\ast+\frac{k_{\bfk_\|}^2}{2m^\ast}|<2t_{z}^\ast.
\end{equation}

%
\wee

%

\end{document}